\documentclass[12pt]{iopart}
\usepackage{times}
\usepackage{iopams}
\usepackage{bm}
\usepackage[dvips]{graphicx}

\begin{document}

\title[Time evolution of spin state of radical ion pair in microwave field]{Time evolution of spin state of
radical ion pair in microwave field: An analytical solution}
\author{S V Anishchik, V N Verkhovlyuk and V A Bagryansky}
\address{Institute of Chemical Kinetics and Combustion, 630090 Novosibirsk, Russia}
\ead{svan@kinetics.nsc.ru, v\_ver@ngs.ru and vbag@kinetics.nsc.ru}
\begin{abstract}
The paper reports an exact solution for the problem of spin evolution of radical ion pair
in static magnetic and resonant microwave field taking into account Zeeman and hyperfine
interactions and spin relaxation. The values of parameters that provide one of the four
possible types of solution are analysed. It is demonstrated that in the absence of spin
relaxation, besides the zero field invariant an invariant at large amplitudes of the
resonant microwave field can be found. The two invariants open the possibility for simple
calculation of microwave pulses to control quantum state of the radical pair. The effect
of relaxation on the invariants is analysed and it is shown that changes in the high
field invariant are induced by phase relaxation.
\end{abstract}
\pacs{03.65.Db, 33.35.+r}
 \submitto{\jpb}
 \maketitle

\section{Introduction}

The problem of controlling elementary chemical reactions, and especially the advances in
the field of quantum informatics \cite{Galindo02} stimulate interest towards controlling
quantum state of microscopic spin systems \cite{Zutic04}. Spin correlated radical pairs
are one of the most intriguing quantum objects. The quantum state of such a pair can be
controlled by applying external magnetic fields \cite{Brocklehurst69,Kniga}. Additional
application of resonant microwave fields substantially widens the possibilities for
manipulating the spins.

\begin{figure}[b]
\centering\includegraphics[width=3.5cm]{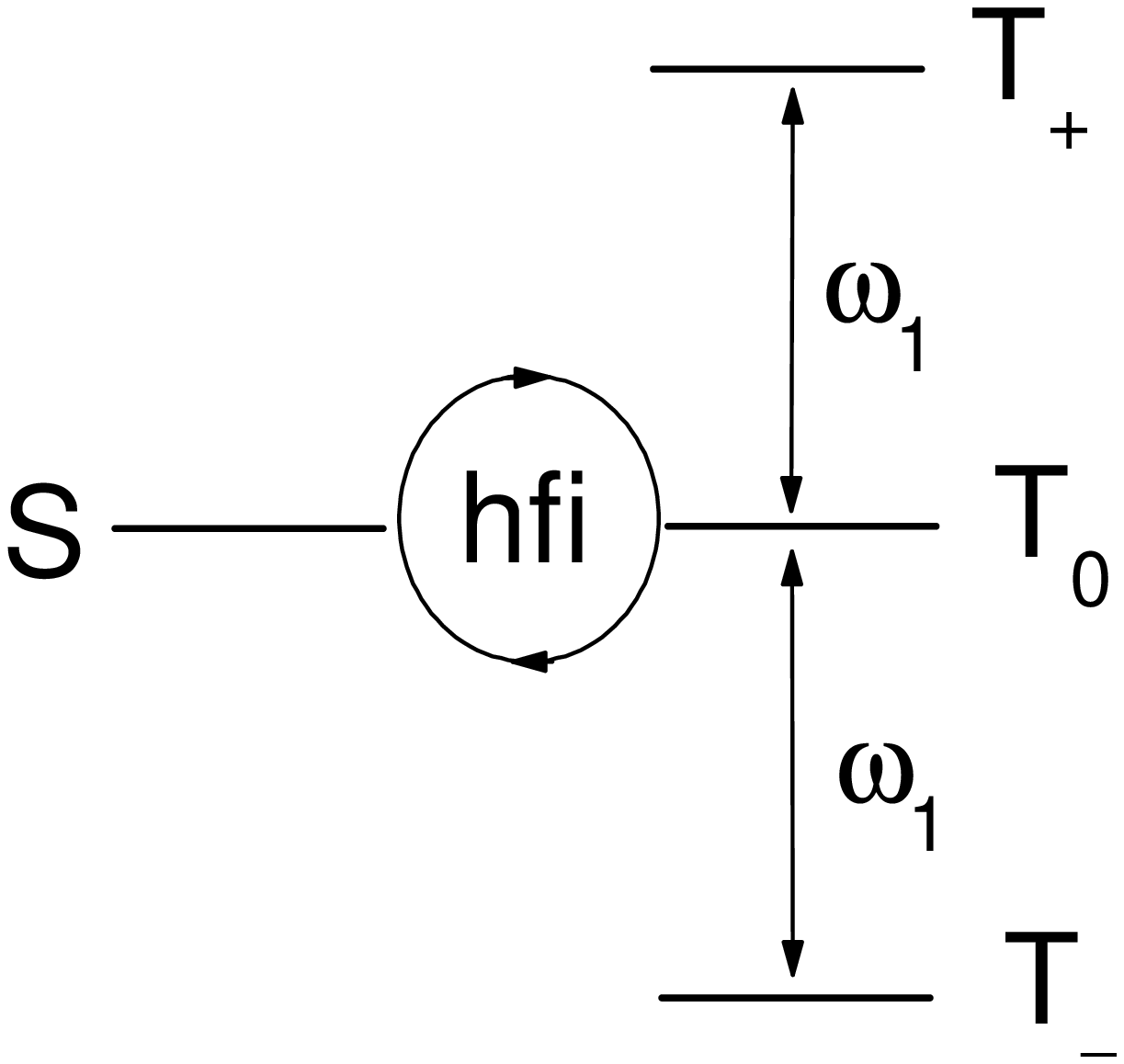} \caption{Spin dynamics of a radical ion
pair} \label{Schema}
\end{figure}

\Fref{Schema} shows the scheme of transitions between spin states of a radical ion pair
induced by hyperfine interaction with magnetic nuclei and resonant microwave field.
Usually the set of singlet state of the pair $|S\rangle$ with zero total spin and three
triplet states $|T_+\rangle$, $|T_0\rangle$ and $|T_-\rangle$ with total spin 1 and its
projection on the $z$ axis equal to $+1$, $0$ and $-1$, respectively, is taken is the
basis. In strong applied magnetic field the degeneracy of the triplet states is lifted by
Zeeman interaction. In the absence of microwave field and relaxation the populations of
$|T_+\rangle$ and $|T_-\rangle$ states, as well as the sum of the populations of
$|T_0\rangle$ and $|S\rangle$ states remain constant. In the presence of hyperfine
interaction the states  $|T_0\rangle$ and $|S\rangle$ are not stationary, and their
populations periodically change with time. This phenomenon is referred to as quantum
beats and is observed in experiment \cite{Klein76,Bizyaev83}. Quantum beats can also be
induced by differences in the $g$-values of the pair partners \cite{Veselov87}.

A resonant microwave field induces transitions between the triplet states, which leads to
changes in the population of singlet state of the pair and can be used to optically
detect ESR spectra of radical ions in liquids \cite{Anisimov79,Trifunac80}. If the rate
of transitions between $|S\rangle$ and $|T_0\rangle$ states is higher than between the
triplet states, periodic changes of the singlet state population with frequency
proportional to amplitude of the microwave field are observed, which is referred to as
quantum oscillations \cite{Saik90,Shkrob95,Anishchik99}. In the opposite limiting case of
strong microwave field, when the rate of transitions between the triplet states exceeds
the rate of transitions between $|S\rangle$ and $|T_0\rangle$, singlet-triplet
transitions are slowed down, and the so-called spin locking is observed. As was
demonstrated \cite{Salikhov93} for pairs formed in singlet initial spin state a strong
microwave field only slows down singlet-triplet transitions but does not completely block
them, and the population of $|T_0\rangle$ state always remains low. This was suggested as
a possible route to control spin state of such a pair. If the microwave field is switched
off at the moment when the population of the singlet state is minimal, most pairs will be
trapped in the $|T_+\rangle$ and $|T_-\rangle$ states. The only way to get into the
singlet state for them is spin-lattice relaxation. A similar idea was exploited to
experimentally substantially prolong lifetime of the radical pair in photosynthetic
reaction center \cite{Dzuba96} by transferring it into its $|T_+\rangle$ and
$|T_-\rangle$ states. Another route to control spin state of the pair was found
theoretically in the work \cite{Anishchik2002}. If a strong resonant microwave field is
rapidly switched on at the moment when spin system is in its $|T_0\rangle$ state, the
pair is completely locked in triplet state and will never become singlet through dynamic
transitions. A possibility of controlling spin state of the pair with very short pulses
of very strong microwave field has also been treated theoretically \cite{Kubarev97b}. The
theory shows that quantum beats in radical pair can be controlled with two (for
singlet-born pairs) or just one (for triplet-born pairs) short microwave pulses.

The already cited works \cite{Salikhov93,Anishchik2002} treated spin dynamics of a
radical pair in static magnetic and resonant microwave fields neglecting relaxation. A
method for solving the problem of spin dynamics taking into account relaxation was
developed in work \cite{Anishchik99}, and approximate solutions in the two limiting cases
of large and small splitting were obtained. A method for numerical solution of this
problem taking into account relaxation and ion-molecular charge transfer reaction was
developed in work \cite{Morozov00}. In the present contribution we provide a general
analytical solution for the problem of spin evolution of a radical ion pair taking into
account spin relaxation and use it to analyse the characteristic features of time
evolution of the pair spin state in microwave field.

\section{Analytical solution}
\subsection{Model}

Let us consider what happens when a non-polar solution is irradiated by ionizing
radiation. As an example we shall use liquid hydrocarbons (RH), in which ionizing
irradiation in the presence of electron (A) and hole acceptors (D) initiates the
processes which in turn lead to formation of several types of singlet spin correlated
pairs $(\,\rm RH{\cdot}^{+}/e^-\,)$, $(\,\rm RH{\cdot}^{+}/A{\cdot}^{-}\,)$, $(\,\rm
D{\cdot}^{+}/e^-\,)$ and $(\,\rm D{\cdot}^{+}/A{\cdot}^{-}\,)$ and their subsequent
recombination \cite{BrocklehurstO1,BrocklehurstO2}. Singlet excited molecules ($\rm
{}^{1}A^{*}$ and $\rm {}^{1}D^{*}$) produced by recombination emit a quantum of light
that is detected by experimental setup. The intensity of the detected luminescence is
thus proportional to probability of forming singlet excited molecules upon recombination.
The effects of magnetic fields on recombination luminescence have as their origin spin
evolution in geminate pairs and dependence of the yield of luminescence on the
multiplicity of excited molecule formed by recombination.

An important advantage of non-polar solutions is that the initial inter-partner distance
in the pairs after ionization (normally $\sim5\div6$~nm) is substantially lower than
Onsager radius ($\sim$30~nm for alkanes at room temperature). The overwhelming majority
of the pairs thus recombine as geminate pairs, i.e., with their sibling counter ion. Upon
ionization of molecule the spin of the ejected electron does not change its state. Since
in the molecule the spins were paired, the initial spin state of the geminate pair is
always singlet. Charge transfer to acceptors and charge recombination also do not change
spin states. This means that the multiplicity of the excited molecule formed by
recombination is determined by the multiplicity of the recombining pair immediately
before recombination. The multiplicity of the excited molecule is thus determined by spin
evolution of the geminate radical ion pair between the moments of ionization and
recombination. Static or oscillating magnetic fields can substantially affect this
evolution. And finally, as excited molecules normally emit light from their singlet
excited state, any effect of magnetic fields on the spin state of the pair is directly
reflected in the intensity of recombination luminescence.

In liquid non-polar solution the partners of the pair formed by irradiation spend almost
the entire period of time from ionization and up to recombination at substantial
distances from each other (tens of nanometers), as their approach to distances about
1--2~nm leads to practically instant recombination. Thus dipole-dipole and exchange
interactions between the pair partners can be safely neglected. Spin evolution of the
pair in this case is driven by interaction of electron spin with spins of nuclei
(hyperfine interaction, HFI), with external magnetic fields, and by spin relaxation
processes.

Spin Hamiltonian for radical pair consisting of radicals $A$ and $D$ in high static
magnetic field $\bm{B}_0$ and microwave filed with magnetic component $\bm{B}_1(t)$ can
be written as follows:
\begin{equation}\label{H2}
\hat{H}=g\beta [\bm{B}_0+\bm{B}_1(t)](\hat{\bm{S}}^A+\hat{\bm{S}}^D)+
\hat{{S}}_{z}^A\sum_{i}a^A_i\hat{{I}}_{zi}^A +\hat{{S}}_z^D\sum_{j}
a^D_j\hat{{I}}_{zj}^D,
\end{equation}
where $\hat{\bm{S}}^A$ and $\hat{\bm{S}}^D$ are spin operators for electrons, and
$\hat{\bm{I}}_{i}^A$ and $\hat{\bm{I}}_{j}^D$ are spin operators for magnetic nuclei in
radicals $A$ and $D$ respectively. The sum runs over the nuclei of radicals. The first
term describes Zeeman interaction of electron spins with external magnetic field, the
second and third terms correspond to isotropic hyperfine interaction of electron spins
with nuclear spins in the two radicals. The $z$ axis is aligned along the direction of
the field $\bm{B}_0$. For simplicity we assume that the partners of the pair have
identical g-values ($g_A=g_D=g$).

As was shown in work \cite{Anishchik99} in the presence of spin relaxation time evolution
of the spin operators for the partners of the radical pair $\hat{\bm{S}}{}^A(t)$ and
$\hat{\bm{S}}{}^D(t)$ is describe with a system of equations in the frame rotating at the
frequency of the applied microwave field that are similar to Bloch equations
\cite{Bloch46,Bloch57}. For operator $\hat{\bm{S}}{}^A(t)$ the system is written as:
\numparts \label{Bl_all}
 \begin{eqnarray}
\label{Bl_e1}
d\hat{{S}}^A_x/dt &=&  -\Delta \omega^A \hat{{S}}^A_y - \hat{{S}}^A_x/T^A_2,\\
 \label{Bl_e2}
d\hat{{S}}^A_y/dt &=& \Delta \omega^A \hat{{S}}^A_x - \omega_1 \hat{{S}}^A_z
- \hat{{S}}^A_y/T^A_2,\hspace{2.7cm}\\
\label{Bl_e3} d\hat{{S}}^A_z/dt &=& \omega_1 \hat{{S}}^A_y - \hat{{S}}^A_z/T^A_1,
 \end{eqnarray}
\endnumparts
where $\omega_1 = g\beta B_1/\hbar$, $T^A_1$ and $T^A_2$ are the times of spin-lattice
and spin-spin relaxation. The $z$ axis is aligned in the direction of the static magnetic
field, and the $x$ axis points along the $B_1$ field. $\Delta\omega^A=\omega^A-\omega_0$
is the detuning of the hyperfine component (HFC) of radical $A$ from resonance. Here
$\omega_0$ is the frequency of the applied microwave field, and \(\omega^A=(g\beta
B_0+\sum a^A_iI^A_i)/\hbar\) is the resonance frequency for radical $A$. Time evolution
for partner $\hat{\bm{S}}{}^D(t)$ is described by similar equations after substituting
$D$ for $A$.

In the absence of interactions between the spins of the pair spin state of the pair can
be described using operators of projection $\hat{{P}}_\psi (t)$ on an arbitrary state
$|\psi \rangle$ written through $\hat{\bm{S}}{}^A(t)$ and $\hat{\bm{S}}{}^D(t)$:
\begin{equation}\label{pr1}
w_\psi (t)=Tr[\hat{{P}}_\psi (t) \hat{{\rho}} (0)],
\end{equation}
where $w_\psi (t)$ is the probability of finding the system in the state $|\psi \rangle$.
Some $\hat{{P}}_\psi (t)$ are given in \ref{appB}.

\subsection{Solution of the system of operator equations}

Equations (\ref{Bl_e1})--(\ref{Bl_e3}) are a system of homogeneous linear differential
equations, and its solution can be represented as an expansion over a complete
orthonormal basis. In our case such a basis is conveniently given by Pauli matrices
\(\hat{\sigma}_x, \hat{\sigma}_y,\) and \(\hat{\sigma}_z \).
\begin{equation}
\label{f4} \hat{{S}}^{A,D}_i(t)=\sum_k{\rm \bf
\Lambda}^{A,D}_{ik}(t)\hat{S}_k(0),~~~i,k=x,y,z,
\end{equation}
where \(\hat{S}_i(0)=\frac{1}{2}\hat{\sigma}_i,~~~ i=x,y,z\).

The system (\ref{Bl_e1})--(\ref{Bl_e3}) was solved by Laplace transform and its inversion
along the lines of the work \cite{Morris94}. A substantial difference from
\cite{Morris94} is that in our case the equations are solved for operators, but this does
not change the basic approach.

Laplace transform of the original equations yields the following system:
\numparts \label{lap_all}
 \begin{eqnarray}
\label{lap_f1} \hspace{1.4cm} (p+1/T_2)\hat{\tilde{S}}_x +
\Delta \omega \hat{\tilde{S}}_y &=&  \hat{S}_x(0),  \\
\label{lap_f2}- \Delta \omega \hat{\tilde{S}}_x + (p+1/T_2)\hat{\tilde{S}}_y +
\omega_1\hat{\tilde{S}}_z &=&   \hat{S}_y(0),\hspace{2.45cm} \\
\label{lap_f3}\hspace{1.3cm}- \omega_1\hat{\tilde{S}}_y + (p+1/T_1)\hat{\tilde{S}}_z &=&
\hat{S}_z(0),
 \end{eqnarray}
\endnumparts
where the tilde sign denotes Laplace transforms of the corresponding spin operators, $p$
is the Laplace parameter. This system is a system of linear algebraic equations with
three unknowns, which has a unique solution when its determinant is nonzero

\begin{equation}\label{det_lap}
{\triangle}(p) = (p+1/T_2)^2(p+1/T_1) + \Delta \omega^2(p+1/T_1) + \omega^2_1(p+1/T_2).
\end{equation}

The solution of the system (\ref{lap_f1})--(\ref{lap_f3}) has the following form:

\begin{equation}\label{l_s_all}
\fl \hat{\tilde{\bm{S}}}=\frac{1}{\triangle (p)}\left(
\begin{array}{lll}
(p+\frac{1}{T_2})(p+ \frac{1}{T_1}) + \omega^2_1&-(p+\frac{1}{T_1}) \Delta\omega&
\Delta\omega \omega_1\\
(p+\frac{1}{T_1})\Delta\omega&(p+\frac{1}{T_1})(p+\frac{1}{T_2})&-(p+\frac{1}{T_2})
\omega_1\\
\Delta\omega \omega_1 &(p+\frac{1}{T_2})\omega_1& \Delta\omega^2 + (p+\frac{1}{T_2})^2
\end{array}
  \right)\hat{\bm{S}}(0)
\end{equation}

To take inverse Laplace transform from expressions (\ref{l_s_all}) we need to factorize
the determinant ${\triangle}(p)$. The determinant is a cubic polynomial with respect to
$p$, and its roots are given by Cardano formula. Depending on whether the discriminant of
the equation
\begin{equation} \label{xi}
\xi=q^2/4+s^3/27
\end{equation}
is positive, negative or zero (two cases of degeneration), four types of solution are
possible. We introduced the following notation:
\begin{equation} \label{q}
 q = 2b^3/27 - bc/3 + d,~~~ s = - b^2/3 + c,~~~b =  2/T_2 + 1/T_1,
\end{equation}
\begin{equation} \label{c}
c = 1/T^2_2 + 2/(T_1T_2) + \Delta\omega^2 + \omega_1^2, ~~ d = 1/(T^2_2T_1) +
\Delta\omega^2/T_1 + \omega_1^2/T_2.
\end{equation}
As can be seen from equations (\ref{l_s_all}), in all cases the following relations hold
true:
\begin{equation} \label{pi_all}
  {\rm \bf \Lambda}_{yx}(t)=-{\rm \bf \Lambda}_{xy}(t),~~~
  {\rm \bf \Lambda}_{zx}(t)={\rm \bf \Lambda}_{xz}(t),~~~
  {\rm \bf \Lambda}_{zy}(t)=-{\rm \bf \Lambda}_{yz}(t).
\end{equation}
{\bfseries In the case $\bm{\xi}\bm{>}\bm{0}$}~ the polynomial ~${\triangle}(p)$ has one
real and two complex conjugate roots (the roots of ${\triangle}(p)$ are $-p_1$ and
$-p_2\pm {\rm i}p_3$): \( {\triangle}(p) = (p + p_1)[(p + p_2)^2 + p^2_3], \) where \(
p_1 = b/3 - (y_{+} + y_{-}),~ p_2 = b/3 +(y_{+} + y_{-})/2,~ p_3 = \sqrt{3}(y_{+} -
y_{-})/2,~ y_\pm =\left(-q/2 \pm \sqrt{\xi}\right)^{1/3}.\) This case corresponds to an
oscillatory solution:
\begin{equation} \label{oscillation_case}
{\rm \bf\Lambda}_{ik}(t)=A_{ik}{\rm e}^{-p_1t}+ B_{ik}{\rm e}^{-p_2t}\cos(p_3t)+
\frac{1}{p_3}C_{ik}{\rm e}^{-p_2t}\sin(p_3t),
\end{equation}
The coefficients were found by procedure similar to the method used in the work
\cite{Morris94}. The expressions for the coefficients are given in \ref{appA}.

The solutions for species type $A$ and type $D$ are similar and differ only in the
corresponding indices ($A$ or $D$) for parameters $\Delta\omega$, $T_1$ and $T_2$.

\ref{appA} does not provide expressions for coefficients ${\rm \bf \Lambda}_{yx}(t)$,
${\rm \bf \Lambda}_{zx}(t)$ and ${\rm \bf \Lambda}_{zy}(t)$, as they can be easily
obtained from formulae (\ref{pi_all}).

{\bfseries In the case of negative discriminant ($\bm{\xi}\bm{<}\bm{0}$)} the polynomial
${\triangle}(p)$ has three real roots: \( {\triangle}(p)=(p+p_1)(p+p_2)(p+p_3), \) where
\( p_1=b/3+2{\rho}\cos(\phi/3),~ p_2=b/3-2{\rho}\cos(\pi/3-\phi/3),~
p_3=b/3-2{\rho}\cos(\pi/3+\phi/3),~ \rho=\pm\sqrt{-s/3},~ \cos(\phi)=q/(2\rho^3)\) and
the sign of $\rho$ must coincide with the sign of $q$. Similar to the first case we
obtain:
\begin{equation}\label{f12}
\nonumber{\rm \bf \Lambda}_{ik}(t)=C^1_{ik}e^{-p_1t}(p_3-p_2)+
C^2_{ik}e^{-p_2t}(p_1-p_3)+C^3_{ik}e^{-p_3t}(p_2-p_1).
\end{equation}
This case corresponds to a overdamped solution (all $p_i>0$).

The expressions for the coefficients $C^l_{ik}$ are given in \ref{appA}. Coefficients
${\rm \bf \Lambda}_{yx}(t)$, ${\rm \bf \Lambda}_{zx}(t)$ and ${\rm \bf \Lambda}_{zy}(t)$
can be found using formulae (\ref{pi_all}).

{\bfseries For $\bm{\xi}\bm{=}\bm{0}$} the determinant ${\triangle}(p)$ has three real
roots, two of which coincide: \( {\triangle}(p)=(p+p_1)(p+p_2)^2, \)  where \(
p_1=b/3-2(-q/2)^{1/3},~~~~p_2=b/3+(-q/2)^{1/3}. \) In this case:
\begin{equation}\label{f13}
{\rm \bf \Lambda}_{ik}(t)=A_{ik}{\rm e}^{-p_1t}+B_{ik}{\rm e}^{-p_2t}+ C_{ik}t{\rm
e}^{-p_2t}.
\end{equation}

{\bf For $\bm{\xi}\bm{=}\bm{0}$~~ and ~~$\bm{q}\bm{=}\bm{0}$} the determinant
${\triangle}(p)$ has three coinciding real roots: \( {\triangle}(p)=(p+p_1)^3, \) where
\( p_1= b/3 =(2/T_2+1/T_1)/3. \)
 In this case:
\begin{equation}\label{f14}
{\rm \bf \Lambda}_{ik}(t)=A_{ik}{\rm e}^{-p_1t}+B_{ik}t{\rm e}^{-p_1t}+
C_{ik}\frac{t^2}{2}{\rm e}^{-p_1t}.
\end{equation}
This case (double degeneration) occurs when ~\( \omega_1=2\sqrt{2}\Delta\omega\) ~and~ \(
(1/T_2-1/T_1)=3\sqrt{3}\Delta\omega. \)

The degenerate cases (\ref{f13}) and (\ref{f14}) never occur in actual calculations, so
we do not provide expressions for their coefficients.

\section{Some peculiarities of spin dymanics}

A convenient basis for spin states of the radical pair is given by Bell functions:
\numparts\label{Bell_all}
\begin{equation}\label{Bell1}
|S\rangle=|\psi^-\rangle=\frac{1}{\sqrt{2}}(|\uparrow\rangle|\downarrow\rangle-|\downarrow\rangle|\uparrow\rangle),
\end{equation}
\begin{equation}\label{Bell2}
|T_x\rangle=|\phi^-\rangle=\frac{1}{\sqrt{2}}(|\uparrow\rangle|\uparrow\rangle-|\downarrow\rangle|\downarrow\rangle),
\end{equation}
\begin{equation}\label{Bell3}
|T_y\rangle=|\phi^+\rangle=\frac{1}{\sqrt{2}}(|\uparrow\rangle|\uparrow\rangle+|\downarrow\rangle|\downarrow\rangle),
\end{equation}
\begin{equation}\label{Bell4}
|T_z\rangle=|\psi^+\rangle=\frac{1}{\sqrt{2}}(|\uparrow\rangle|\downarrow\rangle+|\downarrow\rangle|\uparrow\rangle),
\end{equation}
\endnumparts
where $|\uparrow\rangle$ and $|\downarrow\rangle$ are eigenstates of radical with spin
projection on $z$ axis equal to $+1/2$ and $-1/2$ respectively. In the product the left
function always corresponds to radical $A$ and the right function to radical $D$. The
singlet function $|S\rangle$ has zero total spin, and the three triplet functions
$|T_x\rangle$, $|T_y\rangle$ and $|T_z\rangle$ have total spin equal to 1 and zero
projection on axes $x$, $y$ and $z$ respectively.

Specialists in spin chemistry are more used to working in the basis of the singlet
function $|S\rangle$ and the three triplet functions $|T_0\rangle$, $|T_+\rangle$ and
$|T_-\rangle$ with defined projection of the total spin on the $z$ axis equal to $0$,
$+1$ and $-1$ respectively (it is this notation that was used in \Fref{Schema}). As
equations (\ref{Bell2})--(\ref{Bell4}) show, the Bell functions are expressed from them
as follows:
\[
 |T_z\rangle=|T_0\rangle,~~~
 |T_x\rangle=\frac{1}{\sqrt{2}}(|T_+\rangle-|T_-\rangle),~~~
 |T_y\rangle=\frac{1}{\sqrt{2}}(|T_+\rangle+|T_-\rangle).
\]

The expressions for spin projection operators $\hat{P}_{ii}$ in the basis $|S\rangle$,
$|T_x\rangle$, $|T_y\rangle$, $|T_z\rangle$ is given in \ref{appB}.

Let us introduce notation for density matrix written in this basis
at time zero:
\begin{equation} \label{rho_0}
   \hat{\rho}(0)=\left(\begin{array}{llll} \rho_{ss}&\rho_{sx}&\rho_{sy}&\rho_{sz}\\
                                           \rho_{xs}&\rho_{xx}&\rho_{xy}&\rho_{xz}\\
                                           \rho_{ys}&\rho_{yx}&\rho_{yy}&\rho_{yz}\\
                                           \rho_{zs}&\rho_{zx}&\rho_{zy}&\rho_{zz}\end{array}\right).
\end{equation}

In our calculations we shall also use a particular case when initial state of the pair at
zero time is given by a wave function that is a linear combination of singlet $|S\rangle$
and triplet $|T_z\rangle$ states:
\begin{equation} \label{psi_0}
|\Psi\rangle(0)=\cos\theta\,|S\rangle+\sin\theta\,{\rm
e}^{i\phi}|T_z\rangle.
\end{equation}
In this case $\rho_{ss}=\cos^2{\theta}$,
$\rho_{zz}=\sin^2{\theta}$,
$\rho_{sz}=\frac{1}{2}\sin{2\theta}e^{-i\phi}$,
$\rho_{zs}=\frac{1}{2}\sin{2\theta}e^{i\phi}$, all other
$\rho_{ij}$ are equal to zero.

\subsection{Spin dynamics in the absence of spin relaxation}

In the absence of spin relaxation, i.e. for $1/{T_1}=1/{T_2}=0$, the calculations become
substantially simpler, which allows to get a number of important results analytically. In
this an oscillatory solution always occurs :
\begin{equation}\label{oscillation_case_norel}
{\rm \bf \Lambda}_{ik}(t)=A_{ik}+ B_{ik}\cos(\omega t)+\frac{1}{\omega}C_{ik}\sin(\omega
t),
\end{equation}
where $\omega=\sqrt{\omega_1^2+\Delta\omega^2}$.

\subsubsection{Spin dynamics in the absence of microwave pumping ($\omega_1=0$)}

As it follows from expressions (\ref{pr_s_s})--(\ref{pr_z_z}),
\begin{equation} \label{pr_s_t_z}
 \hat{P}_{ss}+\hat{P}_{zz}=|S\rangle\langle S|+|T_z\rangle\langle
 T_z|=\frac{1}{2}-2\hat{S}_z^A(t)\hat{S}_z^D(t),
\end{equation}
\begin{equation} \label{pr_t_x_t_y}
 \hat{P}_{xx}+\hat{P}_{yy}=|T_x\rangle\langle T_x|+|T_y\rangle\langle T_y|=
 \frac{1}{2}+2\hat{S}_z^A(t)\hat{S}_z^D(t).
\end{equation}

For each radical we can write down:
\begin{equation}
\fl \hat{S}_z(t)=\frac{\Delta\omega\omega_1}{\omega^2}\left(1-\cos(\omega
t)\right)\hat{S}_x(0)+ \frac{\omega_1}{\omega}\sin(\omega t)\hat{S}_y(0)+
\left(\frac{\Delta\omega^2}{\omega^2}+\frac{\omega_1^2}{\omega^2}\cos(\omega
t)\right)\hat{S}_z(0).
\end{equation}

For $\omega_1=0$  and considering (\ref{tr_ro_xx})--(\ref{tr_ro_zy}) we have:
\begin{equation} \label{Tr_zz}
 Tr[\hat{S}^A_z(t)\hat{S}^D_z(t)\rho(0)]=\frac{1}{4}(-\rho_{ss}+\rho_{xx}+\rho_{yy}-\rho_{zz}) ,
\end{equation}
where we used the notation of (\ref{rho_0}).

From (\ref{pr_s_t_z}), (\ref{Tr_zz}) and from the normalization
condition
\begin{equation}\label{norm_rho}
\rho_{ss}+\rho_{xx}+\rho_{yy}+\rho_{zz}=1
\end{equation}
it follows that for $\omega_1=0$ the sum of populations of the
singlet state $\rho_{ss}(t)$ and teh state $|T_z\rangle$
$\rho_{zz}(t)$ does not depend on time:
\begin{equation}\label{zero_mw}
\rho_{ss}(t)+\rho_{zz}(t)=\rho_{ss}+\rho_{zz}.
\end{equation}
This is a quite natural result, since in the absence of microwave pumping and spin
relaxation the only possible transitions in the system are between the states $|S\rangle$
and $|T_z\rangle$ due to hyperfine interaction or differences in the $g$-values of the
pair partners. The two other triplet states $|T_+\rangle$ and $|T_-\rangle$ (that can be
linearly combined to produce the $|T_x\rangle$ and $|T_y\rangle$ states) in these
conditions are stationary, and their population does not change in time due to
substantial Zeeman splitting.

\subsubsection{Spin dynamics in a strong microwave field
($\Delta\omega\! /\omega_1\rightarrow 0$)}

As it follows from expressions (\ref{pr_s_s})--(\ref{pr_z_z}),
\begin{equation} \label{pr_s_t_x}
 \hat{P}_{ss}+\hat{P}_{xx}=|S\rangle\langle S|+|T_x\rangle\langle
 T_x|=\frac{1}{2}-2\hat{S}_x^A(t)\hat{S}_x^D(t),
\end{equation}
\begin{equation} \label{pr_t_y_t_z}
 \hat{P}_{yy}+\hat{P}_{zz}=|T_y\rangle\langle T_y|+|T_z\rangle\langle T_z|=
 \frac{1}{2}+2\hat{S}_x^A(t)\hat{S}_x^D(t).
\end{equation}

Since for each radical
\begin{equation}
\fl
\hat{S}_x(t)=\left(\frac{\omega_1^2}{\omega^2}+\frac{\Delta\omega^2}{\omega^2}\cos(\omega
t) \right) \hat{S}_x(0)- \frac{\Delta\omega}{\omega}\sin(\omega t)\hat{S}_y(0)+
\frac{\Delta\omega\omega_1}{\omega^2}\left(1-\cos(\omega t)\right)\hat{S}_z(0),
\end{equation}
for $\Delta\omega/\omega_1\rightarrow 0$ considering (\ref{tr_ro_xx})--(\ref{tr_ro_zy})
we have:
\begin{equation} \label{Tr_xx}
 Tr[\hat{S}^A_x(t)\hat{S}^D_x(t)\rho(0)]\longrightarrow \frac{1}{4}
 (-\rho_{ss}-\rho_{xx}+\rho_{yy}+\rho_{zz}).
\end{equation}

From (\ref{pr_s_t_x}), (\ref{Tr_xx}) and (\ref{norm_rho}) we obtain that in the limit of
strong microwave field a new invariant appears. The sum of populations of the singlet
$|S\rangle$ and the triplet $|T_x\rangle$ states does not depend on time:
\begin{equation}\label{strong_mw}
\rho_{ss}(t)+\rho_{xx}(t)\approx \rho_{ss}+\rho_{xx}.
\end{equation}
The equality becomes the more accurate the smaller is the ratio $\Delta\omega/\omega_1$.

\subsubsection{Spin dynamics in microwave fields of finite strength}

\begin{figure}[b]
\centering\includegraphics[width=.6\textwidth]{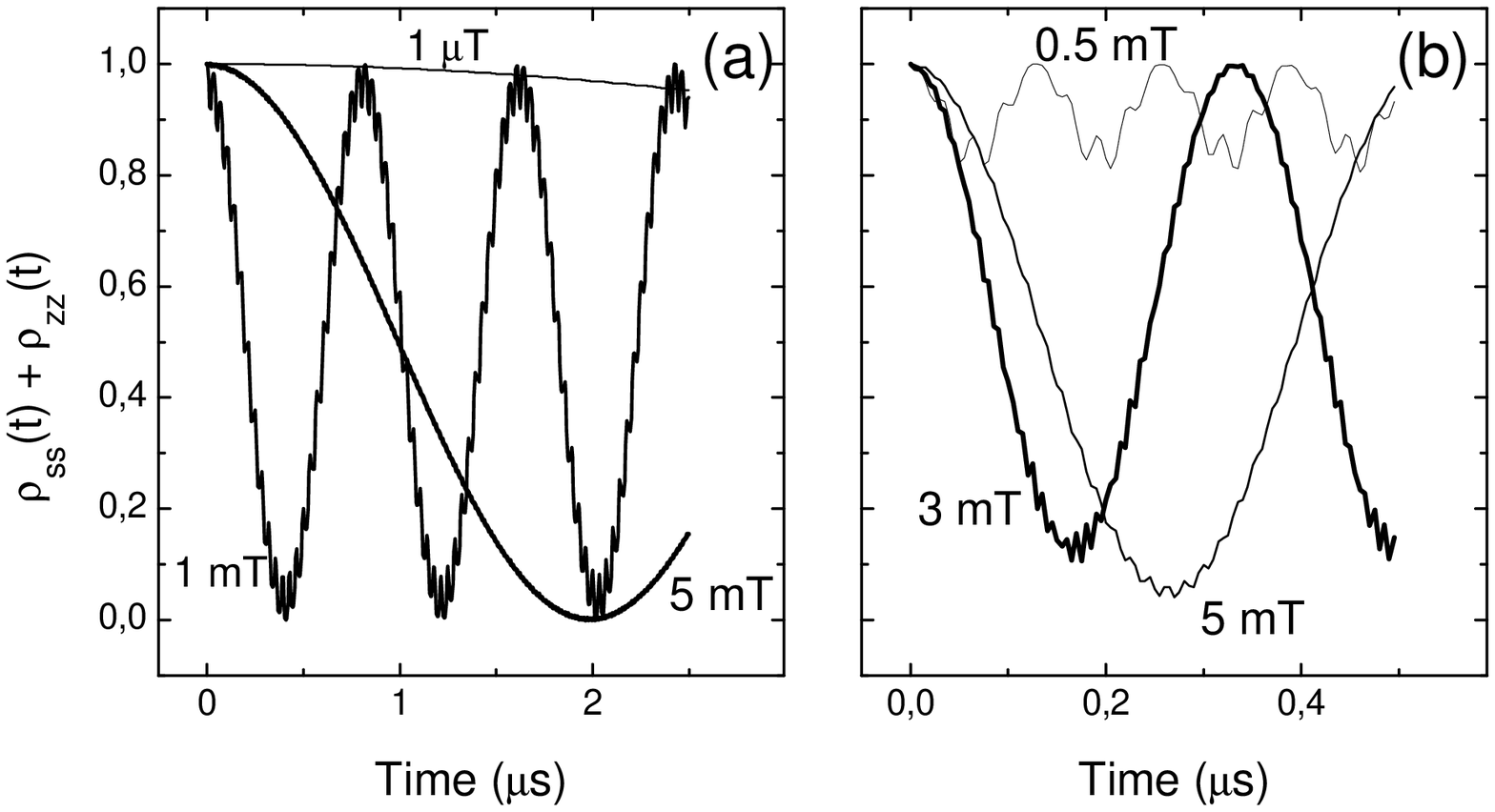}%
\caption{\label{small} Time dependence of the sum of populations of the singlet
$|S\rangle$ and the triplet $|T_z\rangle$ states.
${\Delta}B_A$=0, ${\Delta}B_D$=0.3 mT (a); ${\Delta}B_A$=1~mT,
${\Delta}B_D$=1.3~mT (b). The value of $B_1$ is indicated next to
its corresponding curve}
\end{figure}

The situation in arbitrary microwave fields is more complex than in the limiting cases
discussed above, so we shall consider the general properties of the obtained solutions
using several specific examples as an aid.

\Fref{small}(a) shows time dependencies of the sum of populations of the singlet
$|S\rangle$ and the triplet $|T_z\rangle$ states of the system at different magnitudes of
$B_1$. The line of one radical is precisely in resonance with the applied microwave
field, and the line of the other partner is detuned by 0.3~mT. Initial state of the pair
was taken to be singlet, i.e. $\rho_{ss}=1$, all other $\rho_{ij}=0$. For \Fref{small}
and all next figures are true $\Delta B_{A,D}=\hbar\Delta\omega^{A,D}/(g\beta)$. As the
figure shows, in this case even a weak microwave field is sufficient to completely
destroy invariant (\ref{zero_mw}). The reason for this is transitions between
$|T_z\rangle$ and $|T_y\rangle$ states induced by resonant microwave field. As a
consequence the value of $\rho_{ss}(t)+\rho_{zz}(t)$ periodically changes from zero to
one. The population of the singlet state also changes periodically (see \Fref{ex_s_2}).
This phenomenon is referred to as quantum oscillations and has been observed
experimentally \cite{Saik90,Shkrob95,Anishchik99} and studied theoretically
\cite{Salikhov93,Anishchik99}. The changes in the frequency of oscillations with
increasing amplitude of the microwave field are not monotonous. The frequency first
increases but then, as $B_1$ grows substantially larger than the splitting between the
lines of the two radicals, falls down. The decrease of the frequency is related to the
effect of spin locking.

\Fref{small}(b) shows time dependences $\rho_{ss}(t)+\rho_{zz}(t)$ in the situation of
strong detuning of the microwave frequency from resonance. Time scale is expanded three
five times with respect to \Fref{small}(a). As is clear from the figure, in this case for
not too large amplitudes of the microwave field the frequency of oscillations
substantially increases as compared to the case of resonance with accompanying decrease
in the amplitude of the oscillations.
\begin{figure}[b]
\centering\includegraphics[width=.5\textwidth]{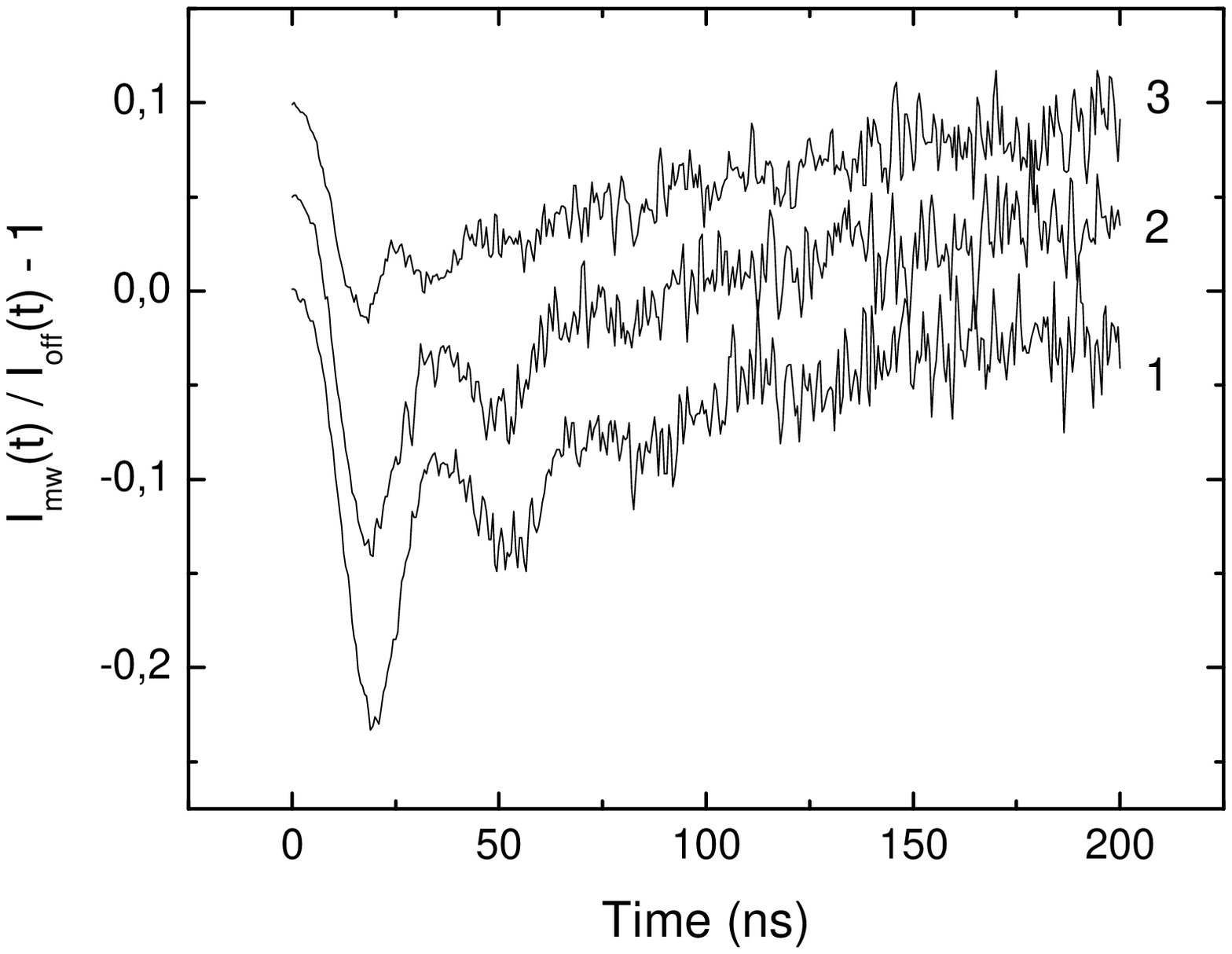}%
\caption{\label{exper} Experimentally observed microwave field effect in the kinetics of
recombination fluorescence from dodecane solution of $10^{-4}$~M ${\rm PTP}-d_{14}$ and
$10^{-2}$~M $\rm C_6F_6$ in the conditions of resonance (1) and with detuning of 0.3~mT
(2) and 1.3~mT (3). $B_1=1$~mT. For convenience the curves 2 and 3 are shifted upwards by
0.05 and 0.1 respectively}
\end{figure}

This effect can be observed experimentally. \Fref{exper} shows experimental results on
monitoring kinetics of recombination fluorescence from liquid alkane solutions in the
absence and in the presence of microwave field. The experimental technique that was used
to get the presented traces was described in details in work \cite{Anishchik99}. The
processes initiated in the sample under the action of a short pulse of X-irradiation are
given by scheme (\ref{reac_all}). Dodecane was used as the solvent (RH),
hexafluorobensene $\rm C_6F_6$ served as the electron acceptor (A), and deuterated {\it
para-}terphenyl ${\rm PTP}-d_{14}$ as the hole acceptor (D). ESR spectrum of ${\rm
PTP}-d_{14}$ radical cation is a single inhomogeneously broadened line with width about
0.1~mT. Radical anion of $\rm C_6F_6$ has a wide resolved spectrum with hyperfine
coupling constant of about 13.5~mT with six equivalent fluorine nuclei
\cite{Anisimov1980}. Time dependence of the intensity of recombination fluorescence
$I(t)$ was sampled using single photon counting technique. The observed light in this
case mostly comes from excited ${\rm PTP}-d_{14}$ molecules. Since recombination kinetics
$f(t)$ does not depend on the multiplicity of the pair, and fluorescence intensity is
proportional to singlet state population of the pair at the moment of recombination
$\rho_{ss}(t)$, we can write that $I(t)\approx f(t)\rho_{ss}(t)$. The equality becomes
more accurate as fluorescence time of the luminophore shortens (about 1~ns for ${\rm
PTP}-d_{14}$) and time resolution of the experimental setup improves (about 3~ns in our
case). In these conditions time resolved microwave field effect is given by the following
expression:
\begin{equation} \label{exp_mw}
I_{mw}(t)/I_{off}(t)-1\approx \rho_{ss}^{mw}(t)/\rho_{ss}^{off}(t)-1,
\end{equation}
where $I_{mw}(t)$ and $\rho_{ss}^{mw}(t)$ are luminescence kinetics and time dependence
of the singlet state population in the presence of microwave field, and $I_{off}(t)$ and
$\rho_{ss}^{off}(t)$ are the same functions in the absence of the field.

\Fref{exper} shows three curves. Curve 1 corresponds to conditions of microwave field
resonance for radical cation of ${\rm PTP}-d_{14}$, curves 2 and 3 --- to detuning from
resonance by 0.3~mT and 1.3~mT respectively. As can be seen from the figure, the
experimental curves are modulated with oscillations. The frequency of oscillations
increases with increasing the detuning, and their amplitude falls down in qualitative
agreement with the results of theoretical calculations shown in \Fref{small}. Because of
rather large HFI constants in radical anion of $\rm C_6F_6$ the populations of
$|S\rangle$ and $|T_z\rangle$ states can be considered equal at any time, and
$\rho_{ss}(t)=\frac{1}{2}[\rho_{ss}(t)+\rho_{zz}(t)]$. Exprimental curves are noisy as
they were not passed through any smoothing routines, and the figure shows that noise
level increases with time. As was shown in work \cite{Anishchik99}, it is proportional to
$t^{3/4}$.

As \Fref{small}(b) shows for large magnitudes of $B_1$ in the case of detuning from
resonance the frequency of oscillations also decreases, but this happens at larger
amplitudes of microwave field than in the conditions of resonance and is accompanied with
increasing amplitude of the oscillations.

\begin{figure}[t]
\centering\includegraphics[width=.8\textwidth]{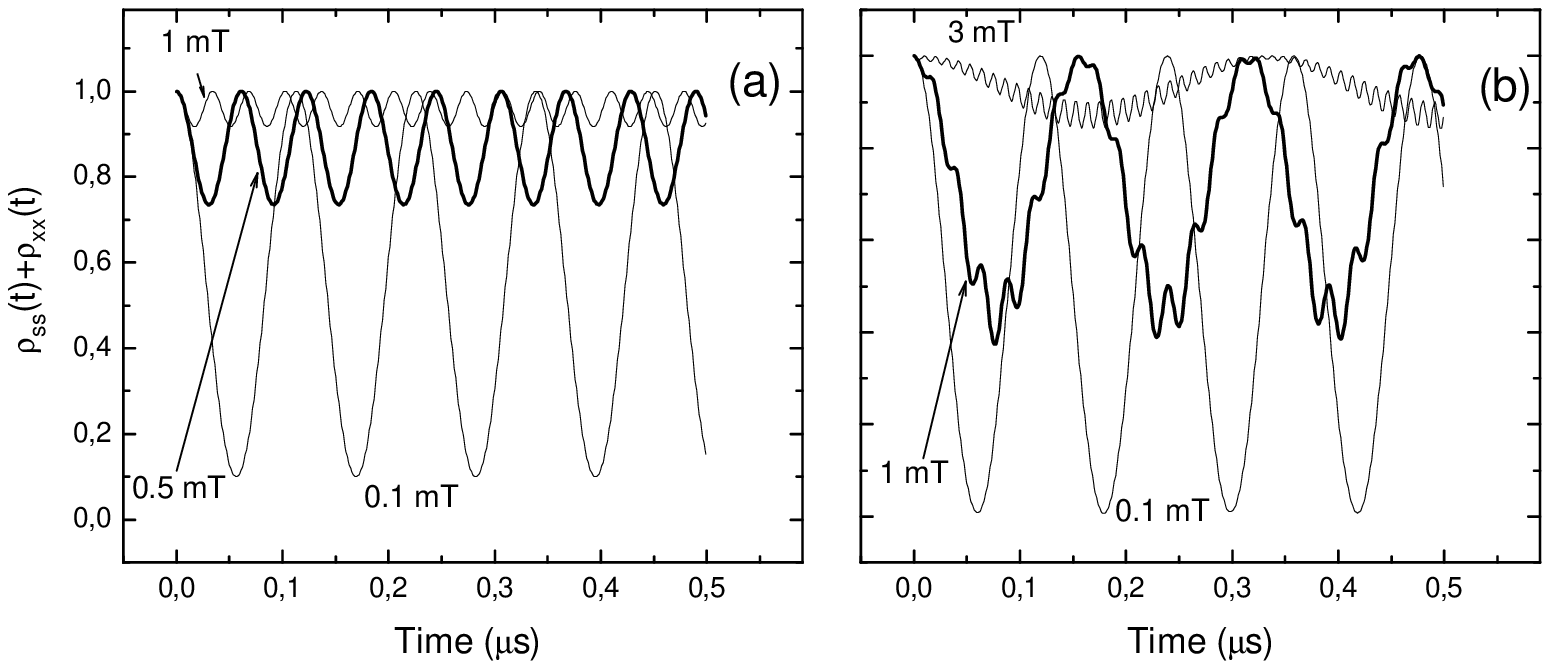}%
\caption{\label{strong} Time dependence of the sum of populations of the singlet
$|S\rangle$ and triplet $|T_x\rangle$ states for different amplitudes of microwave field.
The magnitude of $B_1$ is indicated next to its corresponding curve. ${\Delta}B_A$=0,
${\Delta}B_D$=0.3 mT (a); ${\Delta}B_A$=1~mT, ${\Delta}B_D$=1.3 mT (b)}
\end{figure}

\Fref{strong}(a) shows time dependences of the sum of populations of the singlet
$|S\rangle$ and the triplet $|T_x\rangle$ states at several amplitudes of microwave
field. In the calculations it was assumed that one radical is precisely in resonance with
the applied microwave field, and the line of the second radical is detuned by 0.3~mT. All
calculations were performed for singlet initial state.
\Fref{strong}(b) shows the results of calculations similar to those shown in
\Fref{strong}(a), but with detuning from resonance by 1~mT.

As \Fref{strong} shows, the behaviour of these curves qualitatively differs from the
curves of \Fref{small}. An increase in the amplitude of microwave field leads to
increased frequency and decreased amplitude of the oscillations. Such a behaviour of the
calculated curves reflects the existence of the high field invariant (\ref{strong_mw}).
It can be noted that the accuracy of conserving the value of $\rho_{ss}(t)+\rho_{xx}(t)$
at small splitting is rather high already for microwave field amplitudes easily
accessible in experiment. The magnitudes of $B_1$ amplitudes required for conserving the
high field invariant increase in the case of detuning from resonance (or increasing the
value of $\Delta\omega$).

\begin{figure}[t]
\centering\includegraphics[width=.75\textwidth]{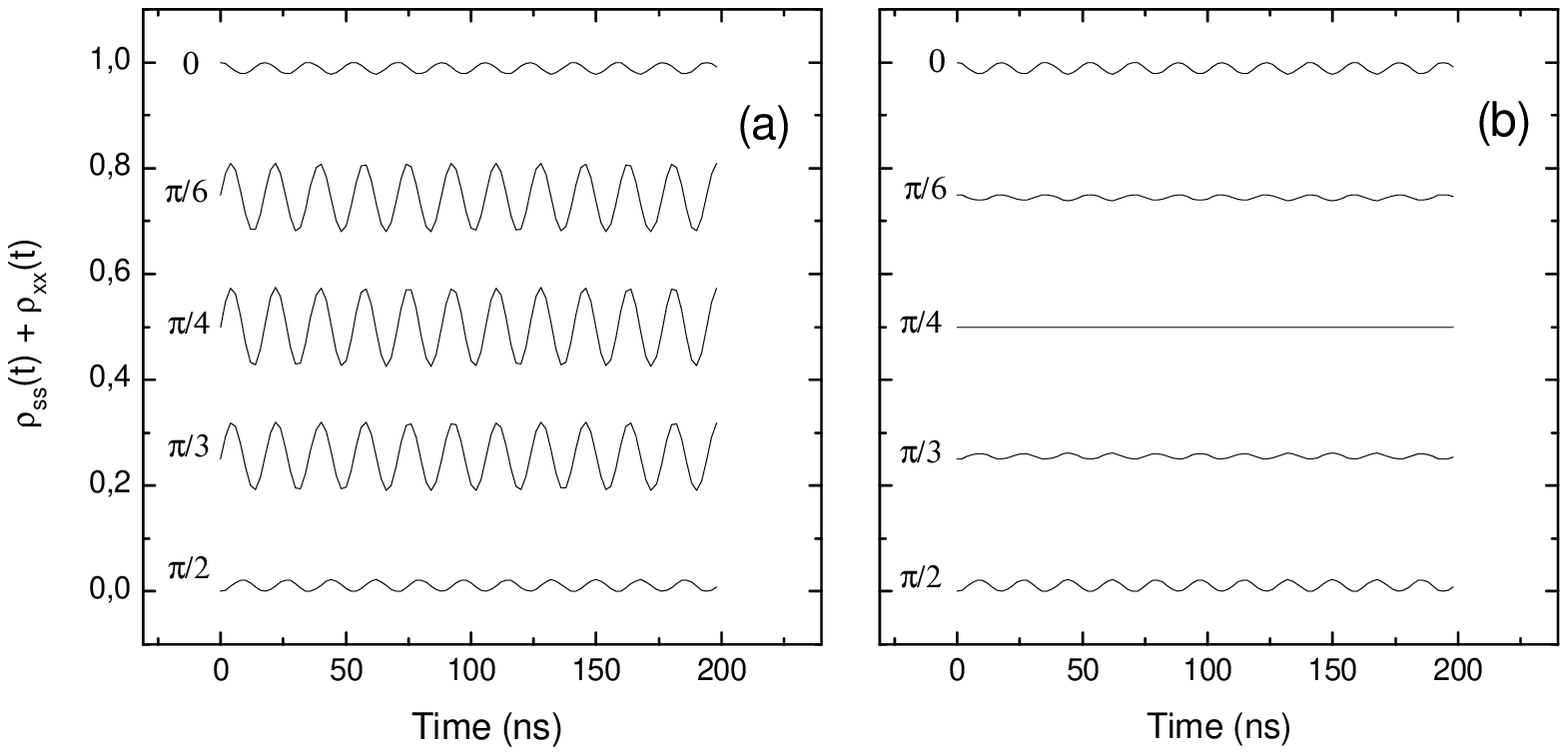}%
\caption{\label{strongmw1} Time dependence of the sum of populations of the singlet
$|S\rangle$ and the triplet $|T_x\rangle$ states for different initial conditions given
by equation (\ref{psi_0}). The values of $\theta$ are indicated next to their
corresponding curves. $\phi=\pi/2$ (a) and 0 (b). ${\triangle}B_A$=0,
${\triangle}B_D$=0.3 mT. $B_1$=2~mT}
\end{figure}

For singlet initial state of the pair and sufficiently large magnitude of $B_1$ the sum
of populations $\rho_{ss}(t)+\rho_{xx}(t)$ oscillate close to one, as can be seen from
\Fref{strong}, which is in good agreement with equation (\ref{strong_mw}). In a more
general case (\ref{psi_0}), when initial singlet state population $\rho_{ss}(0)$ is equal
to $\cos^2 \theta$, the sum of populations $\rho_{ss}(t)+\rho_{xx}(t)$ should oscillate
close to this value $\cos^2 \theta$. The correctness of this conclusion is demonstrated
by the results of calculations given in \Fref{strongmw1}, which shows time dependences of
the sum of populations of the singlet $|S\rangle$ and the triplet $|T_x\rangle$ states of
the system for different initial conditions (\ref{psi_0}). The curves that are shown in
this figure correspond to several sets of values of $\theta$ and $\phi$. As can be
clearly seen, the curves in fact oscillate close to the level of $\cos^2\theta$. The
frequency of these oscillations is indeed close to $\omega_1$, and their amplitude
depends on $\phi$.

\subsection{On possibility to control spin state of the radical ion pair}

The existence of invariants (\ref{zero_mw}) and (\ref{strong_mw}) significantly
simplifies analysis of spin evolution of radical ion pairs in the absence of microwave
field and in strong microwave fields. This opens important possibilities for controlled
impact on the pair to manipulate its spin state.

As was demonstrated in \cite{Anishchik2002}, the existence of the high field invariant
(\ref{strong_mw}) allows in principle to perform ''true spin locking''. Upon rapid
application of a strong microwave field at the moment when the pair is in its triplet
$|T_z\rangle$ state the system is ``locked'' in triplet state and cannot get into singlet
state. At the same time for singlet initial state the system cannot be completely
``locked'', and singlet state population oscillates at a frequency close to
$\Delta\omega_{AD}^2/2\omega_1$, where $\Delta\omega_{AD}=\omega^A-\omega^D$, where
$\omega^A$ and $\omega^D$ are resonance frequencies for radicals $A$ and $D$ respectively
\cite{Salikhov93}. This effect is made absolutely clear by the existence of the high
field invariant (\ref{strong_mw}). For singlet initial state the sum of populations
$\rho_{ss}(t)+\rho_{xx}(t)$ is equal to 1 all the time, and the populations
$\rho_{ss}(t)$ and $\rho_{xx}(t)$ change in antiphase with frequency
$\Delta\omega_{AD}^2/2\omega_1$. The population of the state $|T_z\rangle$ always remains
very close to zero \cite{Salikhov93,Anishchik2002}. From the invariant (\ref{strong_mw})
it follows that the population of the state $|T_y\rangle$ also remains zero. For triplet
initial state $|T_z\rangle$ the sum of populations $\rho_{ss}(t)$ and $\rho_{xx}(t)$
remains zero all the time, which means that populations $\rho_{ss}(t)$ and $\rho_{xx}(t)$
remain zero.

\begin{figure}[t]
\centering\includegraphics[width=.45\textwidth]{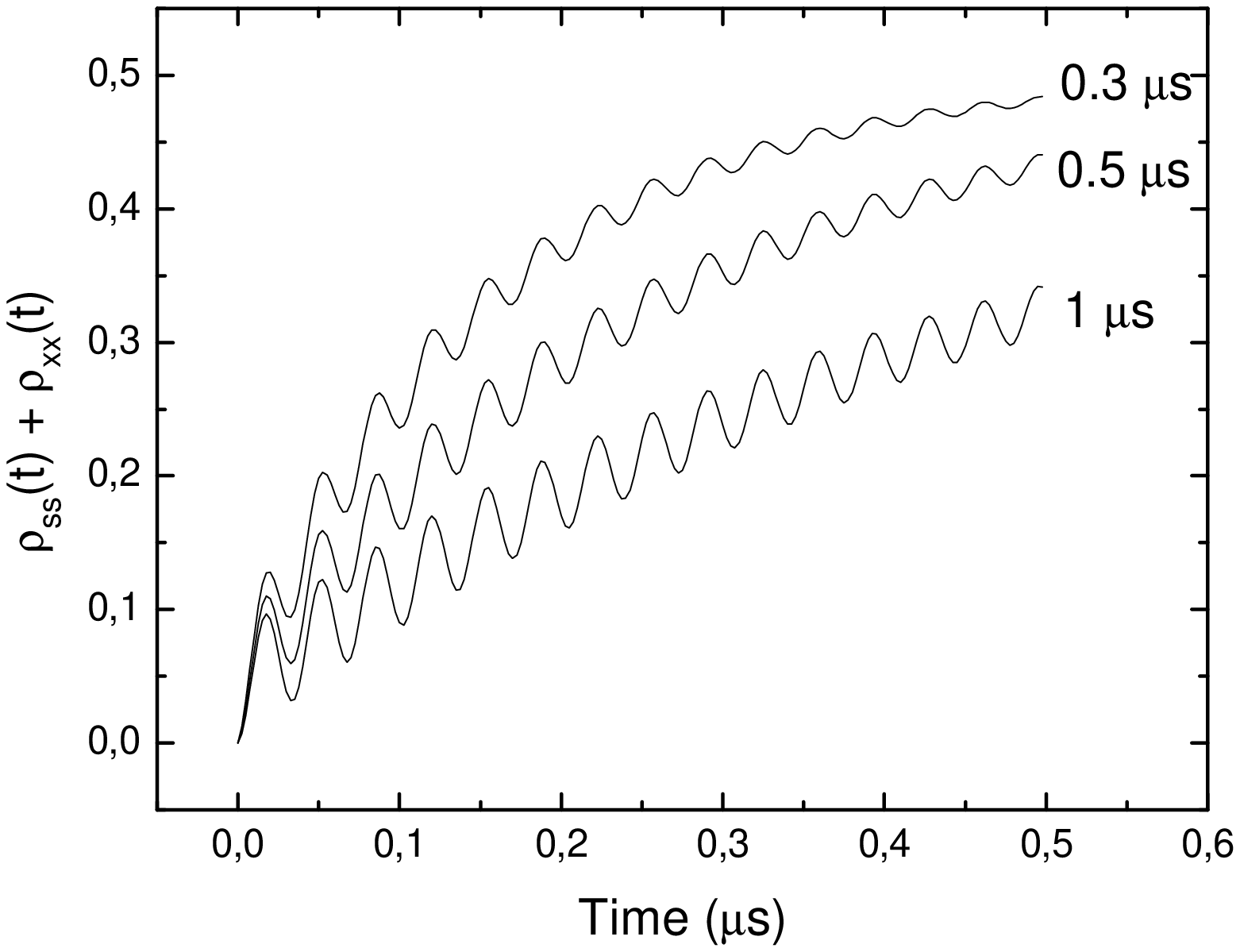}%
\caption{\label{ex_s3_1} Time dependence of the sum of populations of the singlet
$|S\rangle$ and the triplet $|T_x\rangle$ states. Initial state is $|T_z\rangle$.
$B_1$=1~mT, $T_1^A$=$T_1^D$=10~$\mu$s,
 $\Delta B_A$=0, $\Delta B_D$=0.3~mT, $T_2^A$=$T_2^D=T_2$.
 Numerical values of $T_2$ are indicated next to the corresponding curves}
\end{figure}
In radiation processes initial state of the pair is always singlet. However the pair can
be transferred into triplet state $|T_z\rangle$ due to existence of invariant
(\ref{zero_mw}). In the absence of microwave field only the states $|S\rangle$ and
$|T_z\rangle$ get populated. If the frequencies of transitions between them
$\Delta\omega_{AD}$ for different pairs in the sample are multiples of each other, as it
happens when there is an odd number of equivalent magnetic nuclei in the radical ion,
there are moments when all pairs are in their $|T_z\rangle$ states. If a strong microwave
field is applied at this time, the effect of ''true spin locking'' can be observed.

\subsection{The effect of relaxation}

The above considerations are valid only in the absence of spin relaxation, which destroys
spin invariants. The zero microwave field invariant (\ref{zero_mw}) is apparently
destroyed by spin-lattice relaxation. Phase relaxation does not affect this invariant.
This means that expression (\ref{zero_mw}) holds true only for times substantially
smaller than $T_1$. The effect of relaxation on the high field invariant
(\ref{strong_mw}) is much less obvious.
\begin{figure}
\centering\includegraphics[width=.7\textwidth]{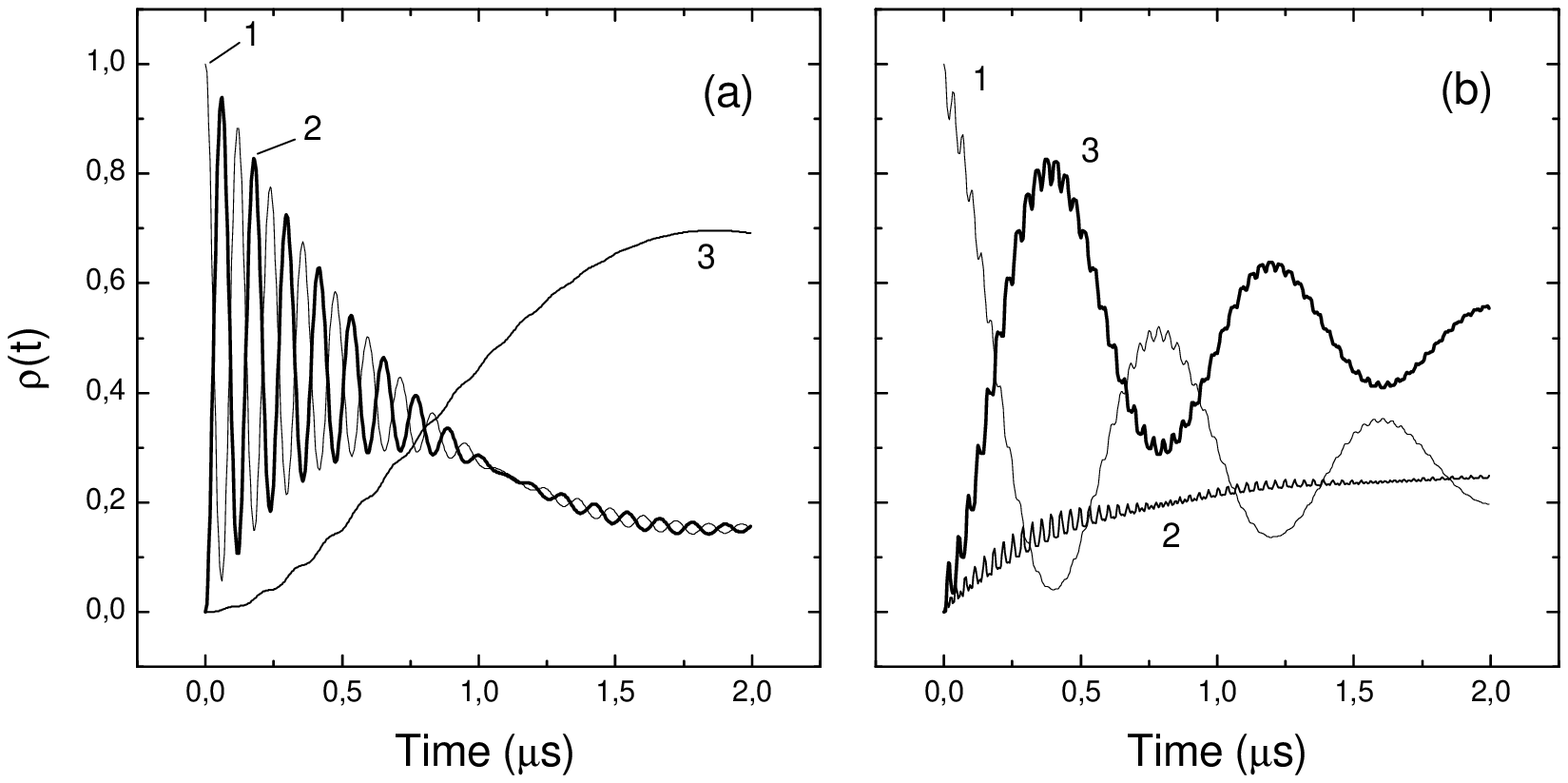}%
\caption{\label{ex_s_2} Time dependence of the populations of the singlet state
$|S\rangle$ (1), triplet state $|T_z\rangle$ (2) and the sum of populations of the
triplet states $|T_x\rangle$ and $|T_y\rangle$ (3) for $T_1^A=T_1^D=1$~s,
$T_2^A=T_2^D=1$~$\mu$s. $\Delta B_A$=0, $\Delta B_D$ = 0.3 mT. $B_1$=10~$\mu$T (a) and
1~mT (b)}
\end{figure}
\Fref{ex_s3_1} shows transformations of time dependence for the sum of populations of the
singlet $|S\rangle$ and the triplet $|T_x\rangle$ states of the pair with changing phase
relaxation time $T_2$. In the absence of relaxation because of invariant
(\ref{strong_mw}) the total population of $\rho_{ss}(t)$ and $\rho_{xx}(t)$ should
oscillate close to zero. However the figure shows that this is no longer the case. The
total population increases with time, and the time scale of this increase is close to the
phase relaxation time $T_2$. We conclude it is phase relaxation that causes the
destruction of the high field invariant (\ref{strong_mw}).

\Fref{ex_s_2} shows time dependences of the populations of different states in the cases
of weak and strong microwave field in the presence of phase relaxation.

The time of spin-lattice relaxation $T_1$ in this model calculation is taken to be very
large, so it has practically no effect on the obtained curves.

As \Fref{ex_s_2} shows, time profile of the populations of different states are
qualitatively different upon application of weak and strong microwave fields. In a weak
field $\rho_{ss}(t)$ and $\rho_{zz}(t)$ oscillate in antiphase with frequency close to
 $\Delta\omega_{AD}$. The amplitude of these oscillations drops because of
phase relaxation, and $\rho_{ss}(t)$ and $\rho_{zz}(t)$ asymptotically approach each
other. The sum of populations $\rho_{xx}(t)+\rho_{yy}(t)$ in this case is a more smooth
curve, which is periodically modulated with frequency close to $\omega_1$.

In a strong microwave field we see the manifestations of spin locking. Singlet state
population $\rho_{ss}(t)$ oscillates at a frequency about $\Delta\omega$ similar to the
weak field case, but the amplitude of these oscillations is rather small. Oscillations of
substantial amplitude occur at a frequency about $\Delta\omega_{AD}^2/2\omega_1$. This is
also the frequency, at which changes the sum of populations $\rho_{xx}(t)+\rho_{yy}(t)$.
The population of the triplet state $|T_z\rangle$ increases with the time of phase
relaxation $T_2$.

\section{Conclusions}

First of all we note that all these results were obtained in the case of fixed nuclear
configurations in both partners of the pair. In the presence of the reaction of
ion-molecular charge transfer this condition is no longer observed, but this apparently
does not change the zero microwave field invariant (\ref{zero_mw}). The high microwave
field invariant (\ref{strong_mw}) will also remain valid if the condition $\Delta\omega\!
/\omega_1\ll 1$ is not violated in the course of the reaction.

Let us formulate the main rules for spin evolution of a radical pair that follow from the
existence of invariants (\ref{zero_mw}) and (\ref{strong_mw}).

\begin{itemize}
\item[~(i)] In the absence of microwave pumping the sum of populations of the states
$|S\rangle$ and $|T_z\rangle$ (or $|S\rangle$ and $|T_0\rangle$ in different notation)
remains constant at times shorter than $T_1$. The populations of these states oscillate
at a frequency  $\omega^A-\omega^D$, where $\omega^A$ and $\omega^D$ are resonance
frequencies for radicals $A$ and $D$ respectively. It also naturally follows that the sum
of populations of the states $|T_x\rangle$ and $|T_y\rangle$ (or $|T_+\rangle$ and
$|T_-\rangle$) does not change as well. Invariant (\ref{zero_mw}) is destroyed by
spin-lattice relaxation. Even weak resonant microwave field also destroys invariant
(\ref{zero_mw}). The sum of populations $\rho_{ss}(t)+\rho_{zz}(t)$ starts oscillating at
a frequency close to $\omega_1$ (quantum oscillations induced by microwave field).

\item[(ii)] In a strong resonant microwave field remain constant the sum of populations
of the states $|S\rangle$ and $|T_x\rangle$, as well as $|T_y\rangle$ and $|T_z\rangle$.
The consequence of this is the effect of spin locking. the high field invariant
(\ref{strong_mw}) is rather accurately conserved already for not very large magnitudes of
$B_1$ quite accessible in experiment. The magnitude of $B_1$ required to maintain the
invariant (\ref{strong_mw}) increases with detuning from resonance. The high field
invariant (\ref{strong_mw}) is destroyed by phase relaxation.

\end{itemize}

The two invariants that we found suggest a way of controlling transitions between spin
states of a radical pair. One possible variant of controlling quantum spin state would be
a rapid switching of a strong microwave field on and off. In this case the pair will
alternatively be in the conditions when one of the invariants is conserved. Another
possible control option may be a rapid change in the phase of the microwave field by
$\pi/2$. In this case the axes $x$ and $y$ are exchanged, which correspondingly changes
the populations of the $|T_x\rangle$ and $|T_y\rangle$ states.

\ack

 The work was supported by RFBR (grant 05-03-32801) and grant ``Support
of leading scientific schools'' No 84.2003.3.

\appendix

\section{Coefficients in the expressions for different types of solutions of equations
for spin dynamics \label{appA}}

\noindent{\em Case $\xi > 0$ (for oscillatory solution){\protect\vspace{1pt}}}

\[
A_{xx}=(\alpha\beta+\omega^2_1)/\gamma ,~ A_{yy}=\alpha\beta/\gamma
,~A_{zz}=(\beta^2+\Delta\omega^2)/\gamma ,
\]
\[
A_{xy}=-\alpha\Delta\omega/\gamma ,~ A_{xz}=\omega_1\Delta\omega/\gamma ,~
A_{yz}=-\omega_1\beta/\gamma ,
\]
\[
 B_{xx}=1-A_{xx},~~  B_{yy}=1-A_{yy},~~  B_{zz}=1-A_{zz};
 \]
\[
 B_{xy}=-A_{xy},~~~~~  B_{xz}=-A_{xz},~~~~~  B_{yz}=-A_{yz};
 \]
\begin{displaymath}
C_{xx}=(\alpha\beta+\omega^2_1)(p_1-p_2)/\gamma +(\alpha+\beta+p_1-p_2),
\end{displaymath}
\begin{displaymath}
C_{yy}=\alpha\beta(p_1-p_2)/\gamma +(\alpha+\beta+p_1-p_2),
\end{displaymath}
\begin{displaymath}
C_{zz}=(\beta^2+\Delta\omega^2)(p_1-p_2)/\gamma + (2\beta+p_1-p_2),
\end{displaymath}
\begin{displaymath}
C_{xy}=-\Delta\omega[\alpha(p_1-p_2)+\gamma]/\gamma
,~C_{xz}=\omega_1\Delta\omega(p_1-p_2)/\gamma ,
\end{displaymath}
\begin{displaymath}
C_{yz}=-\omega_1(\gamma+\beta(p_1-p_2))/\gamma ,
\end{displaymath}
\noindent where
 \[
 \alpha=1/T_1-p_1,~\beta=1/T_2-p_1,~\gamma=(p_2-p_1)^2+p_3^2.
 \]

{\protect\vspace{1pt}}\noindent{\em Case $\xi < 0$ (for overdamped
solution){\protect\vspace{1pt}}}

\begin{displaymath}
C^l_{xx}=\frac{1}{\textstyle
\eta}\{p^2_l-(\frac{1}{T_1}+\frac{1}{T_2})p_l+
\omega^2_1+\frac{1}{T_1T_2}\},
\end{displaymath}
\begin{displaymath}
C^l_{yy}=\frac{1}{\textstyle
\eta}\{p^2_l-(\frac{1}{T_1}+\frac{1}{T_2})p_l+
\frac{1}{T_1T_2}\},
\end{displaymath}
\begin{displaymath}
C^l_{zz}=\frac{1}{\textstyle \eta}\{p^2_l-\frac{2}{T_2}p_l+
\Delta\omega^2+\frac{1}{T^2_2}\},
\end{displaymath}
\begin{displaymath}
C^l_{xy}=\frac{\Delta\textstyle \omega}{\textstyle \eta}
(p_l-\frac{1}{T_1}),~C^l_{xz}=\frac{1}{\textstyle \eta}\Delta \omega
\omega_1,~C^l_{yz}=\frac{\textstyle \omega_1}{\textstyle \eta}(p_l-\frac{1}{T_2}),
\end{displaymath}

\noindent where $l=1,2,3$, ~~~$ \eta=(p_1-p_2)(p_2-p_3)(p_3-p_1)$.

\section{The values of parameters that lead to certain types of solution for equations
of spin dynamics\label{appC}}

To analyse the different cases of spin dynamics it would be very useful to have simple
rules for determining the type of solution for the equations with the given set of
parameters. As it turned out, these rules can be easily formulated, which will be done in
this section.

As has been shown earlier, the exact analytical solution for the equations of spin
dynamics can belong to one of four types: oscillatory, overdamped, and two degenerate,
depending in the first place on the value of $\xi$.

To simplify further treatment let us introduce the new parameter:
\begin{displaymath}
z=1/T_2-1/T_1.
\end{displaymath}
Since $T_2\leqslant T_1$ in all cases, $z\geqslant 0$.

It turns out that relaxation times $T_1$ and $T_2$ enter the expressions for $q$, $s$ and
$\xi$ (\ref{xi})-(\ref{c}) only as the $z$ parameter:
\begin{displaymath}
q=\frac{1}{3}(\omega_1^2-\Delta\omega^2)z-\frac{2}{27}z^3,~~~
s=-\frac{1}{3}z^2+\Delta\omega^2+\omega_1^2,
\end{displaymath}
\begin{equation}\label{bagr1}
\nonumber\xi=\frac{1}{108}\left[4\Delta\omega^2z^4+(8\Delta\omega^4-20\Delta\omega^2\omega_1^2-\omega_1^4)z^2+
4(\Delta\omega^2+\omega_1^2)^3 \right].
\end{equation}
This means that the type of the sought solution depends only on the differences in the
rates of spin-spin and spin-lattice relaxation, but not on the absolute values of the two
rates.

As can be seen from expression (\ref{bagr1}), $\xi$ is a quadratic function with respect
to $z^2$, with branches of the parabola going upwards. This means that in the absence of
real roots $\xi$ is positive for all values of $z$. If the roots exist $\xi$ is negative
for values of $z^2$ in-between the roots and is positive for $z^2$ outside this range.

From $\xi=0$ it follows:
\begin{equation}\label{bagr2}
\nonumber z_{1,2}^2=\frac{1}{8\Delta\omega^2}\left(\omega_1^4+20\Delta\omega^2\omega_1^2-
8\Delta\omega^4 \pm \sqrt{\omega_1^2(\omega_1^2-8\Delta\omega^2)^3} \right).
\end{equation}
The solutions exist for $\omega_1\geqslant2\sqrt{2}\Delta\omega$. It can be easily shown
that both roots are positive.

From this we conclude that $\xi>0$ either for $\omega_1<2\sqrt{2}\Delta\omega$, or for
$\omega_1>2\sqrt{2}\Delta\omega$ when $z<z_2$ or $z>z_1$ ($z_1$ corresponds to choice of
``+'' in (\ref{bagr2}), and $z_2$ to the choice of ``$-$'' respectively). In this case we
obtain a solution of the oscillatory type.

$\xi<0$  for $\omega_1>2\sqrt{2}\Delta\omega$ in the range $z_2<z<z_1$. This set of
parameters produces a solution of the overdamped type.

For $\xi=0$, when either $z=z_1$ or $z=z_2$ we have a solution of the degenerate type.

Finally, when simultaneously $\xi=0$ and $q=0$ the case of double degeneracy takes place.
It can be easily shown that this occurs for $\omega_1=2\sqrt{2}\Delta\omega$ and
$z=3\sqrt{3}\Delta\omega$.

Especially interesting is the behaviour of the solution in the vicinity of resonance for
one of the radical ions, that is, when \( \Delta\omega/\omega_1\rightarrow 0\). Small
parameter expansion of (\ref{bagr2}) then yields:
\begin{displaymath}
z_1\approx \omega_1^2/(2\Delta\omega)\rightarrow\infty,~~~ z_2\approx2\omega_1.
\end{displaymath}
Thus close to the resonance line an oscillatory solution will be obtained for \(
\omega_1>\frac{1}{2}(1/T_2-1/T_1), \)  and an overdamped solution for
 \( \omega_1<\frac{1}{2}(1/T_2-1/T_1). \)

\section{Spin operators\label{appB}}

 Below we express projection operators of the basis states in the representation
$|S\rangle$, $|{\rm T_x}\rangle$, $|T_y\rangle$, $|T_z\rangle$ in terms of spin operators
of the pair partners:
\begin{equation} \label{pr_s_s}
 \hat{P}_{ss}=|S\rangle\langle S|=\frac{1}{4}-\hat{S}_x^A\hat{S}_x^D-\hat{S}_y^A\hat{S}_y^D-\hat{S}_z^A\hat{S}_z^D,
\end{equation}
\begin{equation} \label{pr_x_x}
 \hat{P}_{xx}=|T_x\rangle\langle T_x|=\frac{1}{4}-\hat{S}_x^A\hat{S}_x^D+\hat{S}_y^A\hat{S}_y^D+\hat{S}_z^A\hat{S}_z^D,
\end{equation}
\begin{equation} \label{pr_y_y}
 \hat{P}_{yy}=|T_y\rangle\langle T_y|=\frac{1}{4}+\hat{S}_x^A\hat{S}_x^D-\hat{S}_y^A\hat{S}_y^D+\hat{S}_z^A\hat{S}_z^D,
\end{equation}
\begin{equation} \label{pr_z_z}
 \hat{P}_{zz}=|T_z\rangle\langle T_z|=\frac{1}{4}+\hat{S}_x^A\hat{S}_x^D+\hat{S}_y^A\hat{S}_y^D-\hat{S}_z^A\hat{S}_z^D.
\end{equation}

Let us express the trace of the product of spin operators by initial density matrix in
the notation of (\ref{rho_0}):
\begin{equation} \label{tr_ro_xx}
Tr[\hat{S}_x^A(0)\hat{S}_x^D(0)\hat{\rho}(0)]=\frac{1}{4}(-\rho_{ss}-\rho_{xx}+\rho_{yy}+\rho_{zz}),
\end{equation}
\begin{equation}
Tr[\hat{S}_y^A(0)\hat{S}_y^D(0)\hat{\rho}(0)]=\frac{1}{4}(-\rho_{ss}+\rho_{xx}-\rho_{yy}+\rho_{zz}),
\end{equation}
\begin{equation}
Tr[\hat{S}_z^A(0)\hat{S}_z^D(0)\hat{\rho}(0)]=\frac{1}{4}(-\rho_{ss}+\rho_{xx}+\rho_{yy}-\rho_{zz}),
\end{equation}
\begin{equation}
Tr[\hat{S}_x^A(0)\hat{S}_y^D(0)\hat{\rho}(0)]=\frac{i}{4}(\rho_{zs}-\rho_{yx}+\rho_{xy}-\rho_{sz}),
\end{equation}
\begin{equation}
Tr[\hat{S}_x^A(0)\hat{S}_z^D(0)\hat{\rho}(0)]=\frac{1}{4}(-\rho_{ys}+\rho_{zx}-\rho_{sy}+\rho_{xz}),
\end{equation}
\begin{equation}
Tr[\hat{S}_y^A(0)\hat{S}_x^D(0)\hat{\rho}(0)]=\frac{i}{4}(-\rho_{zs}-\rho_{yx}+\rho_{xy}+\rho_{sz}),
\end{equation}
\begin{equation}
Tr[\hat{S}_y^A(0)\hat{S}_z^D(0)\hat{\rho}(0)]=\frac{i}{4}(-\rho_{xs}+\rho_{sx}-\rho_{zy}+\rho_{yz}),
\end{equation}
\begin{equation}
Tr[\hat{S}_z^A(0)\hat{S}_x^D(0)\hat{\rho}(0)]=\frac{1}{4}(\rho_{ys}+\rho_{zx}+\rho_{sy}+\rho_{xz}),
\end{equation}
\begin{equation}\label{tr_ro_zy}
Tr[\hat{S}_z^A(0)\hat{S}_y^D(0)\hat{\rho}(0)]=\frac{i}{4}(\rho_{xs}-\rho_{sx}-\rho_{zy}+\rho_{yz}).
\end{equation}

\end{document}